\documentclass[preprint,aps,showpacs]{revtex4}
\usepackage{amsfonts}
\usepackage{graphics}
\usepackage{graphicx}
\usepackage{epsf}

\newcommand{\be}{\begin{equation}}
\newcommand{\ee}{\end{equation}}
\newcommand{\bea}{\begin{eqnarray}}
\newcommand{\eea}{\end{eqnarray}}
\newcommand{\sech}{{\rm sech}}

\begin{document}
\title{Note on the Gauss-Bonnet braneworld scenario}

\author{D. Bazeia$^1$, A. Lob\~{a}o$^2$, L. Losano$^1$, R. Menezes$^{3,4}$, A. Yu. Petrov$^1$}
\affiliation{$^1$Departamento de F\'{\i}sica, Universidade Federal da Para\'{\i}ba, 58051-970, Jo\~ao Pessoa, Para\'{\i}ba, Brazil}
\affiliation{$^2$Escola T\'ecnica de Sa\'ude de Cajazeiras, Universidade Federal de Campina Grande, 58900-000 Cajazeiras, PB, Brazil}
\affiliation{$^3$Departamento de Ci\^encias Exatas, Universidade Federal da Para\'{\i}ba, 58297-000 Rio Tinto, PB, Brazil}
\affiliation{$^4$Departamento de F\'\i sica, Universidade Federal de Campina Grande, 58109-970 Campina Grande, PB, Brazil}

\pacs{04.50.-h, 11.25.-w}
\date{\today}

\begin{abstract}
In this work we deal with the presence of braneworld solutions in a five-dimensional space-time with a single extra spatial dimension of infinite extent. The braneworld scenario is built under the presence of a single real scalar field, and we modify the gravity sector to include generic function of the Gauss-Bonnet term. We study several specific models, and we construct exact braneworld solutions, in particular for including the Gauss-Bonnet term at first and second order power. As an interesting result, we show that the brane tends to split for a specific modification in the gravity sector, in the presence of non constant Gauss-Bonnet term.
\end{abstract}

\maketitle 

The study of gravity is certainly one of the most fundamental research lines in the modern theoretical high energy physics. The problem of explaining the cosmic acceleration discovered in \cite{Riess}, together with search for a consistent quantum gravity model, called a great deal of attention to the study of modified theories of gravity. The crucial role was played by the paper \cite{RS} where the idea of noncompact extra dimensions, which gave rise to wide application of the brane concepts in gravity and cosmology, has been introduced.

Following the brane concept, our four-dimensional space-time is treated as a brane embedded to some higher-dimensional space-time, that is, the bulk where the gravitational field can propagate, whereas the fields of fundamental interactions are confined to the brane \cite{bw}. In the standard scenario, the higher-dimensional gravity was supposed to be described by the usual Einstein-Hilbert action. However, a very natural development of the idea consisted in modifying of the gravity Lagrangian implemented through the replacement of the scalar curvature $R$ by some other scalar depending on the gravitational fields and involving its higher derivatives. Within the quantum field theory, the interest to such a modification of the gravity has been arisen by the seminal paper \cite{Stelle} where the higher-derivative extensions of gravity were considered as a way to achieve a renormalizable gravity theory. Within the classical gravity, study of the $f(R)$ models was developed as an aim for a possible explanation of a cosmic acceleration \cite{fr}. Many issues related to the different aspects of the $f(R)$ gravity, especially the exact solutions in this theory, have been studied in \cite{fraspects}. Within the brane context, the $f(R)$ gravity has been applied for the first time in \cite{frbrane}.

However, the $f(R)$ gravity does not include all possibilities for extensions of gravity. One of the most studied extensions is the Lovelock gravity, such that the gravity action is now represented as a series, with its lowest (zero) order being a cosmological term, the first order is the scalar curvature, and the second one is the Gauss-Bonnet term \cite{Lovelock}.
As a result, the Gauss-Bonnet gravity naturally emerges, in which the gravity Lagrangian is a sum of the usual Einstein-Hilbert Lagrangian with the Gauss-Bonnet scalar or its function. It is well-known that in the four-dimensional space-time, the Gauss-Bonnet term is topological. Some aspects of the Gauss-Bonnet gravity have been considered earlier in \cite{gaussbonnet}.  In \cite{GBonne1}, this theory has been applied within the cosmic acceleration context. It should be mentioned, that the higher-order contractions of the Riemann curvature tensor, and hence higher orders in the Gauss-Bonnet term, naturally arise in the low-energy limit of the string theory \cite{Myers}.

Within this paper, we are going to study the Gauss-Bonnet gravity within the brane context. Some earlier studies in this direction have been performed in \cite{GBbrane,Mota}, where, however, only the simplest version of the Gauss-Bonnet gravity has been considered, that is, only the Gauss-Bonnet term $G$ itself is added to the gravity action. At the same time, actually more sophisticated manners to include the Gauss-Bonnet term are carried out, implying thus in the theories of $f(G)$ gravity and $f(R,G)$ gravity which are intensively discussed within the cosmic acceleration problem, for a review see f.e. \cite{FelSu} and references therein. It is worth mentioning that $f(G)$ gravity has been tested within the cosmological observations in \cite{Davis}, where the restrictions on the numerical parameters of the action have been obtained.  Besides of the cosmological studies, the stability issues also have been discussed for the Gauss-Bonnet gravity \cite{Edelstein}, where, moreover, it was shown that just for the case $f(G)=G$, the Gauss-Bonnet gravity displays the instabilities, which certainly implies a suggestion that namely a theory described by a more sophisticated function $f(G)$ can be free of instabilities. Therefore, it seems natural to apply these more generalized models also within the braneworld context.

Our aim will consist principally in applying the first-order formalism which has manifested itself as a very useful tool for solving the nonlinear equations of motion, especially modified Einstein equations \cite{frbrane}. Using this method, we will find some new exact solution for the Gauss-Bonnet brane. The exact solutions are then used to investigate the profile of the energy density, searching in particular for the brane splitting feature, as it has appeared before in \cite{a,b} and more recently in \cite{blmps}. In the recent work \cite{blmps}, one noticed that the splitting appears due to  modification in the geometry, with the inclusion of the $R^2$ term in the Einstein-Hilbert action. In the current work we modify the geometry including the Gauss-Bonnet term, so we believe that the splitting is not excluded, being an effect unseen in previous investigations.

We start with the following action describing the Gauss-Bonnet brane (cf. \cite{Mota}):
\bea
S=\int d^4xdy\sqrt{-g}\left(\frac{1}{2}R+\frac{1}{2}f(G)+{\cal L}_s\right).
\eea
Here $y$ is the extra coordinate, ${\cal L}_s$ is the source Lagrange density. As usual, $G$ stands for the Gauss-Bonnet term, which is given by
\bea
G=R^2-4R^{\mu\nu}R_{\mu\nu}+R_{\mu\nu\lambda\rho}R^{\mu\nu\lambda\rho}.
\eea
In the present work, we take the line element in the form
\bea
\label{metric}
ds^2=e^{2A(y)}\eta_{ab}dx^adx^b-dy^2,
\eea 
with $A$ representing the warp function. This stands for $AdS_5$ geometry, and here we consider the source as a scalar field, specified by
\bea
{\cal L}_s=\frac{1}{2}g_{\mu\nu}\partial^{\mu}\phi\partial^{\nu}\phi-V(\phi).
\eea
The modified Einstein equation looks like \cite{Mota}
\bea
\label{eqmod}
G_{\mu\nu}&=&2T_{\mu\nu}+\frac{1}{2}g_{\mu\nu}f(G)-2F(G)RR_{\mu\nu}+4F(G)R_{\mu}^{\lambda}R_{\nu\lambda}-2F(G)R_{\mu\lambda\rho\sigma}R_{\nu}^{\phantom{\nu}\lambda\rho\sigma}\nonumber\\&-&
4F(G)R_{\mu\rho\sigma\nu}R^{\rho\sigma}+2R\nabla_{\mu}\nabla_{\nu}F(G)-2Rg_{\mu\nu}\nabla^2F(G)-4R_{\mu}^{\rho}\nabla_{\nu}\nabla_{\rho}F(G)\nonumber\\&-& 4R_{\nu}^{\rho}\nabla_{\mu}\nabla_{\rho}F(G)+4R_{\mu\nu}\nabla^2F(G)+4g_{\mu\nu}R^{\lambda\rho}\nabla_{\lambda}\nabla_{\rho}F(G)-4R_{\mu\nu\lambda\rho}\nabla^{\lambda}\nabla^{\rho}F(G)\nonumber\\
&\equiv& 2T_{\mu\nu}+H_{\mu\nu}.
\eea
Here we are using $F(G)=\frac{df(G)}{dG}$, $T_{\mu\nu}$ is the energy-momentum tensor of the source field, and $G_{\mu\nu}=R_{\mu\nu}-\frac{1}{2}Rg_{\mu\nu}$ is the usual Einstein tensor. Since $G$ is a scalar depending only on $y\,(=x_4)$, by symmetry the last term $4R_{\mu\nu\lambda\rho}\nabla^{\lambda}\nabla^{\rho}F(G)=0$; unfortunately, however, this does not simplify the expression too much.  For the sake of simplicity we choose in this paper $f(G)=aG^n$, with $n\geq 1$. We use Latin indices $a,b,c,d=0...3$, and Greek ones $\mu\nu,\lambda\rho=0...4$. Also, $A^{\prime}=\frac{dA}{dy}$.
It is clear that for $f(G)=0$ the usual equations of motion \cite{b} are restored.

Direct calculations show that for the braneworld metric (\ref{metric}) one has
\bea
R_{abcd}&=&e^{4A}A^{\prime 2}(\eta_{ac}\eta_{bd}-\eta_{ad}\eta_{bc}); \quad\, R_{4b4d}=-e^{2A}(A^{\prime\prime}+A^{\prime 2})\eta_{bd};\nonumber\\
R_{bd}&=&e^{2A}\eta_{bd}(A^{\prime\prime}+4A^{\prime 2}); \quad\, R_{44}=-4(A^{\prime\prime}+A^{\prime 2}); \quad\, R=8A^{\prime\prime}+20A^{\prime 2};\nonumber\\
R_{\mu\nu}R^{\mu\nu}&=&80A^{\prime 4}+20A^{\prime\prime2}+64A^{\prime\prime}A^{\prime2};\quad\,
R_{\mu\nu\lambda\rho}R^{\mu\nu\lambda\rho}=40A^{\prime4}+16A^{\prime\prime2}+32A^{\prime\prime}A^{\prime2};\nonumber\\
G&=&24[4A^{\prime\prime}A^{\prime2}+5A^{\prime4}];\nonumber\\
T_{ab}&=&\eta_{ab}(\frac{1}{2}\phi^{\prime2}+V(\phi))e^{2A}; \quad\, T_{44}=\frac{1}{2}\phi^{\prime2}-V(\phi).
\eea
For the usual Einstein tensor $G_{\mu\nu}$ one finds in this metric
\bea
G_{ab}=-3(A^{\prime\prime}+2A^{\prime2})\eta_{ab}e^{2A}; \quad\, G_{44}=6A^{\prime2}; \quad\, G_{a4}=0.
\eea

Let us now look for some possibilities to solve the equations (\ref{eqmod}).
The most complicated terms are the derivatives of $f(G)$. As a first step, one can verify that the expression for the Gauss-Bonnet term is correct. Indeed, let us suggest that the metric in the ${\cal D}$-dimensional bulk space-time looks like (\ref{metric}), but with $a,b=0...{\cal D}-2$. Repeating all calculations, one gets
\bea
G=({\cal D}-1)({\cal D}-2)({\cal D}-3)[{\cal D}A^{\prime4} +4A^{\prime\prime}A^{\prime2}].
\eea
At ${\cal D}=5$, one indeed yields $G=120A^{\prime4}+96A^{\prime\prime}A^{\prime2}$. At ${\cal D}=4$, one has $G_4=24A^{\prime4} +24A^{\prime\prime}A^{\prime2}$; in this case, the contribution to the action, that is, $\sqrt{|g|}G_4=24e^{3A}(A^{\prime4} +A^{\prime\prime}A^{\prime2})=8(e^{3A}A^{\prime3})^{\prime}$ (recall that $\sqrt{|g|}=e^{({\cal D}-1)A}$) is a total derivative, thus confirming the known fact that the Gauss-Bonnet Lagrangian is a total derivative in four dimensions. Also, it vanishes in lower dimensions.

The next step is to suggest that the Gauss-Bonnet term $G$ is constant, i.e.
\bea
\label{gb}
\frac{G}{24}=4A^{\prime\prime}A^{\prime2}+5A^{\prime4}=5B.
\eea
In this case, for the positive $B=b^2$, one gets
\bea
y+C=\frac{4}{5}\int\frac{dA^{\prime}\;A^{\prime2}}{b^2-A^{\prime4}},
\eea
and for the negative $B=-b^2$, it gives
\bea
y+C=-\frac{4}{5}\int\frac{dA^{\prime}\;A^{\prime2}}{b^2+A^{\prime4}}.
\eea
In principle, the explicit forms of $A(y)$ can be found for both these factors, although it is ugly: for $B=b^2$, one has
\bea
y+C=-\frac{4}{5}\Big(
\frac{1}{4\sqrt{b}}\ln\Biggl|\frac{A^{\prime}+\sqrt{b}}{A^{\prime}-\sqrt{b}}\Biggr|+\frac{1}{2\sqrt{b}}\arctan\frac{A^{\prime}}{\sqrt{b}}
\Big),
\eea
and for $B=-b^2$, one has an even more complicated expression. However, the result evidently exists.
So, let us write the equations of motion for this case:
\bea
\label{eqmod1}
G_{\mu\nu}&=&2T_{\mu\nu}+\frac{1}{2}g_{\mu\nu}f(G)-2F(G)RR_{\mu\nu}+4F(G)R_{\mu}^{\lambda}R_{\nu\lambda}-2F(G)R_{\mu\lambda\rho\sigma}R_{\nu}^{\phantom{\nu}\lambda\rho\sigma}\nonumber\\&-&
4F(G)R_{\mu\rho\sigma\nu}R^{\rho\sigma}.
\eea
Here $f(G)$ and $F(G)$ are constants. So, we have for the cases $\mu\nu=ab$ and $\mu\nu=44$, respectively:
\bea
\label{firstsys}
-3(A^{\prime\prime}+2A^{\prime2})&=&(\phi^{\prime2}+2V(\phi))+\frac{1}{2}f(G)-F(G)(48A^{\prime4}+
36A^{\prime\prime}A^{\prime2});\\
6A^{\prime2}&=&\phi^{\prime2}-2V(\phi)-\frac{1}{2}f(G)+F(G)(48 A^{\prime4}+48A^{\prime\prime}A^{\prime2}),\nonumber
\eea
or, as is the same after replacement of the first equation by the sum of the two equations,
\bea
\label{syseq}
-3A^{\prime\prime}&=&2\phi^{\prime2}+12 F(G)A^{\prime\prime}A^{\prime2};\nonumber\\
6A^{\prime2}&=&\phi^{\prime2}-2V(\phi)-\frac{1}{2}f(G)+48F(G)(A^{\prime4}+A^{\prime\prime}A^{\prime2}).
\eea
So, we can use some choice for the warp factor consistent with the constant $G$ and arrive at the system on $\phi$ and $V(\phi)$.
We note that within this methodology, the potential $V(\phi)$, instead of being introduced as a basic characteristic of the theory, turns out to be a variable to be determined. However, the assumptions of this method, imposing restrictions on the potential \cite{b,frbrane}, are necessary to employ the power of the first-order formalism allowing to obtain exact solutions for a certain class of potentials while, in a generic case, the solutions can only be found numerically. Therefore, within this approach, we solve a kind of {\it inverse problem}, allowing to find the potential for the known warp factor. Moreover, we demonstrate how the explicit form of the potential can be found.

Let us substitute $G=120D$, with $D$ being a constant. Also, we suppose that if $A'=A'(y)$, one can use $y=y(A')$ as well, therefore one has
\bea
\phi^{\prime}=5\left(\frac{D-A^{\prime4})}{4A^{\prime2}}\right)\frac{d\phi}{dA^{\prime}}.
\eea
We substitute $A^{\prime}\equiv z$. Thus, our system (\ref{syseq}) implies
\bea
-15\left(\frac{D-z^4}{z^2}\right)&=&50\left(\frac{d\phi}{dz}\right)^2\left(\frac{D-z^4}{4z^2}\right)^2+60\tilde{F}(D)(D-z^4);\nonumber\\
6z^2&=&25\left(\frac{D-z^4}{4z^2}\right)^2\left(\frac{d\phi}{dz}\right)^2-2V(\phi)-\frac{1}{2}\tilde{f}(D)+48\tilde{F}(D)(5D-z^4)\,,
\eea
where $\tilde{F}(D)=F(G(D))=F(120D)$, similarly, $\tilde{f}(D)=f(120D)$.
Resolving these equations, we find
\bea
\left(\frac{d\phi}{dz}\right)^2&=&\frac{15}{50}(1+4\tilde{F}(D)z^2)\frac{4z^2}{z^4-D};\nonumber\\
V(z(\phi))&=&\frac{1}{2}\Big[\frac{z^4-D}{4z^2}\frac{15}{2}(1+4F(D)z^2)-6z^2-\frac{1}{2}\tilde{f}(D)+\nonumber\\&+&
48\tilde{F}(D)(5D-z^4)\Big].
\eea
So, in principle, the field and the potential are found in terms of the auxiliary variable $z$.

At the same time, we can abandon the restriction of the constant $G$. In this case, the $H_{ab}$ and $H_{44}$ terms take the form
\bea
H_{ab}&=&\eta_{ab}e^{2 A}\Big[\frac12f(G)-12F(G)A^{\prime2}\left(4 A^{\prime2}+3 A^{\prime\prime}\right)+24  F^{\prime}(G)A^{\prime}\left(A^{\prime2}+A^{\prime\prime}\right)\nonumber\\ &+&
4F^{\prime\prime}(G) \left(2A^{\prime2}- A^{\prime\prime}\right)\Big];\nonumber\\
H_{44}&=&-\frac12f(G)+48 F(G) A^{\prime 2}\left( A^{\prime \prime}+ A^{\prime 2}\right)+16F^{\prime} (G)A^{\prime } \left(A^{\prime \prime}-2A^{\prime 2}\right).
\eea
We consider these equations for the particular case $f(G)=\alpha G^n$ which yields
\bea\label{simplif}
H_{ab}&=&\eta_{ab}e^{2 A}\Big\{\frac12\alpha G^n-12\alpha n A^{\prime2}\left(4 A^{\prime2}+3 A^{\prime\prime}\right)G^{n-1}\nonumber\\
&&+4\alpha n(n-1) \Big[6G^{\prime}  A^{\prime}\left(A^{\prime2}+A^{\prime\prime}\right)+\left(2A^{\prime2}- A^{\prime\prime}\right)G^{\prime\prime}\Big]G^{n-2}\nonumber\\
&&+4\alpha n(n-1)(n-2) G^{\prime 2}\left(2A^{\prime2}- A^{\prime\prime}\right)G^{n-3}\Big\};\nonumber\\
H_{44}&=&-\frac12\alpha G^n+48 \alpha n  A^{\prime 2}\left( A^{\prime \prime}+ A^{\prime 2}\right)G^{n-1}+16\alpha n(n-1) G^{\prime} A^{\prime } \left(A^{\prime \prime}-2A^{\prime 2}\right)G^{n-2}.
\eea
Summing up the equations for $T$ (the $T$ is an object defined from the relation $T_{ab}=\eta_{ab}e^{2A}T$) and $T_{44}$, we find
\begin{equation}
\phi^{\prime 2}=-\frac{3}{2}A^{\prime\prime}-\frac{1}{8}e^{-2A}\eta^{ab}H_{ab}-\frac{1}{2}H_{44}\,.
\end{equation}
Therefore, one has
\bea
\label{eqphi}
\phi^{\prime 2}&=&-\frac32A^{\prime\prime}-6\alpha n A^{\prime2}A^{\prime\prime}G^{n-1}\nonumber\\
&&-2\alpha n(n-1) \Big[\left(2A^{\prime2}- A^{\prime\prime}\right)G^{\prime\prime}-2 A^{\prime}\left(A^{\prime2}-5A^{\prime\prime}\right)G^{\prime} \Big]G^{n-2}\nonumber\\
&&-2\alpha n(n-1)(n-2) G^{\prime 2}\left(2A^{\prime2}-A^{\prime\prime}\right)G^{n-3}.
\eea

It is instructive to give here the explicit expressions for the potential and energy density.
To do it, we note that the modified Einstein equations will look like
\bea
-3(A^{\prime\prime}+2A^{\prime\,2})=(\phi^{\prime\,2}+2V)+H;\nonumber\\
6A^{\prime\,2}=\left(\phi^{\prime\,2}-2 V\right)+H_{44},
\eea
where $H$ can be read off from Eq. (\ref{simplif}).
Subtracting the second equation from the first one, we find the potential:
\bea
V(\phi)=\frac{1}{4}\left(-3(A^{\prime\prime}+4A^{\prime\,2})-H+H_{44}\right),
\eea
whose explicit form is
\bea\label{Vphi}
V(\phi)&=&-\frac{3}{4}A^{\prime\prime}-3A^{\prime\,2}-\frac14\alpha G^n+3\alpha n A^{\prime2}\left(8 A^{\prime2}+7 A^{\prime\prime}\right)G^{n-1}\nonumber\\
&&-\alpha n(n-1) \Big(  A^{\prime}\left(14A^{\prime2}+2A^{\prime\prime}\right)G^{\prime}+\left(2A^{\prime2}- A^{\prime\prime}\right)G^{\prime\prime}\Big)G^{n-2}\nonumber\\&&-\alpha n(n-1)(n-2) G^{\prime 2}\left(2A^{\prime\,2}- A^{\prime\prime}\right)G^{n-3}.
\eea
The energy density in our case is given by the $T_{00}$ component of the energy-momentum tensor of the matter
\bea\label{eneden}
\rho=-e^{2A}{\cal L}=e^{2A}\left(\frac{1}{2}\phi^{\prime 2}+V(\phi)\right)\,.
\eea

\begin{figure}[t]
\begin{center}
\includegraphics[scale=0.4]{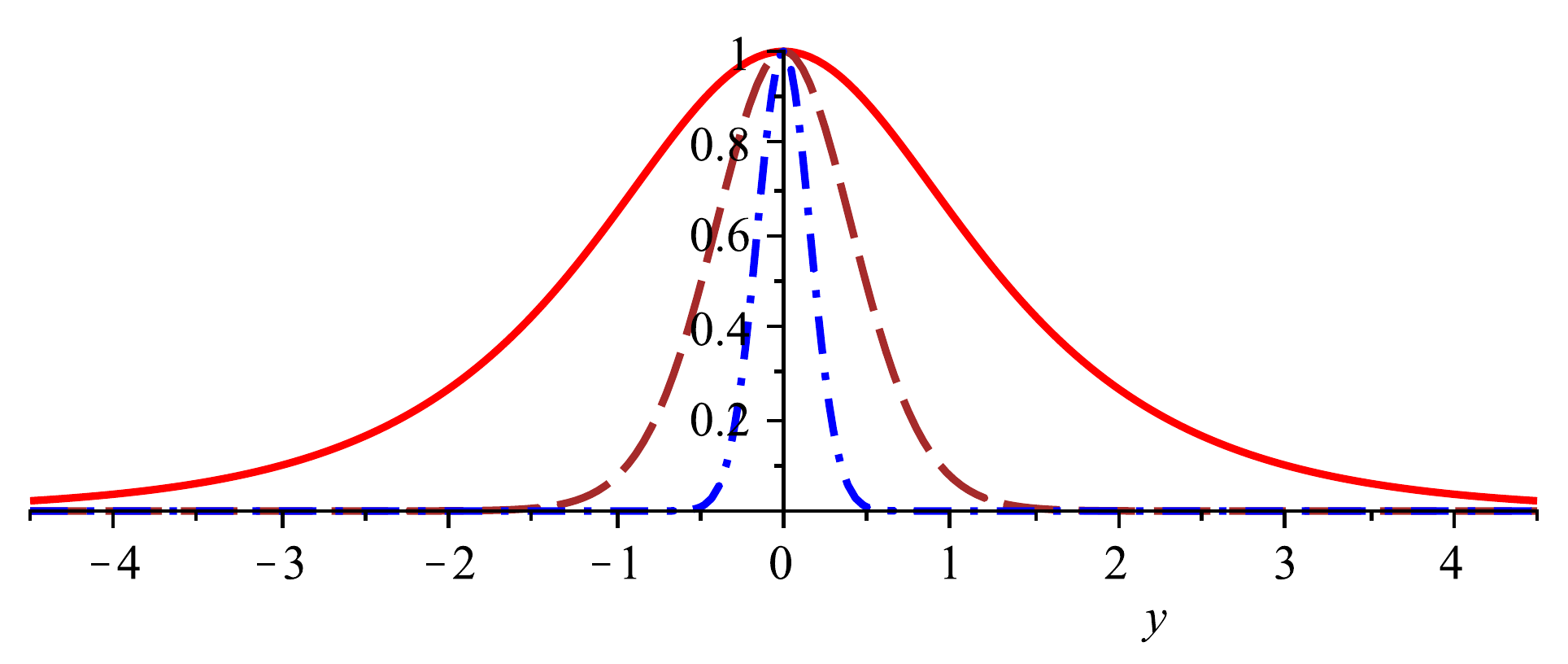}
\end{center}
\vspace{-0.5cm}
\caption{\small{The warp factor (\ref{ansatz}), depicted for $B=0.5$ (red, solid line), $B=3.0,$ (brown, dashed line) and $B=20.0$ (blue, dotted-dashed line).\label{fig1}}}
\end{figure}

Let us now consider some specific cases.

\paragraph{First example.}
For $n=1$ and $\alpha=-1/4$, we find that
\begin{equation}
\label{eq10}
\phi^{\prime 2}=-\frac32A^{\prime\prime}+\frac32A^{\prime 2}A^{\prime\prime}\,.
\end{equation} 
We choose a standard ansatz (which at large $|y|$ tends to $A(y)=-B|y|$):
\begin{equation}
\label{ansatz}
A(y)=B \ln[\sech(y)]\,,
\end{equation}
where $B>0$. We plot the warp function $e^{2A(y)}$ for the above warp function in Fig.~\ref{fig1}. We note that the warp factor becomes more and more localized around $y=0$, as $B$ increases to larger and larger values.

We use the Eq.~(\ref{eq10}) to get
\begin{equation}
\phi^{\prime 2}=\frac{3B}{2} \left(1-B^2\right)\sech^2(y)+\frac{3B^3}{2} \sech^4(y)\,.
\end{equation}
Moreover, in the case $B=1$, we find
\begin{equation}
\phi^{\prime}=\pm \sqrt{\frac32}\sech^2(y)\,.
\end{equation}
The solution of this equation is kink-like:
\begin{equation}
\phi(y)=\pm \sqrt{\frac32}\tanh(y)\,.
\end{equation}

\paragraph{Second example.} We use the same ansatz for the warp function, as in Eq.~(\ref{ansatz}), but now we consider $n=2$. We get
\bea
\phi^{\prime 2}=-\frac32A^{\prime\prime}-4\alpha\Big(3 A^{\prime2}A^{\prime\prime}G+\left(2A^{\prime2}- A^{\prime\prime}\right)G^{\prime\prime}-2 A^{\prime}\left(A^{\prime2}-5A^{\prime\prime}\right)G^{\prime} \Big)\,.
\eea

Using the expression for the Gauss-Bonnet term (\ref{gb}), we arrive at
\bea
\phi^{\prime 2}&=&-\frac32A^{\prime\prime}+96\alpha\Big(25  A^{\prime6}  A^{\prime\prime}-32A^{\prime5}A^{\prime\prime\prime}-8A^{\prime4} A^{\prime\prime\prime\prime}+8 A^{\prime\prime4}- 316 A^{\prime4}A^{\prime\prime2}\nonumber\\
&&-36A^{\prime2} A^{\prime\prime 3}-68A^{\prime3} A^{\prime\prime}A^{\prime\prime\prime}+4A^{\prime2} A^{\prime\prime} A^{\prime\prime\prime\prime} +24A^{\prime} A^{\prime\prime 2} A^{\prime\prime\prime} \Big)\,.
\eea
Using the warp function (\ref{ansatz}), we find
\bea
\label{eqant}
\phi^{\prime2}&=&\frac{3}{2}B\left(1+64\alpha B^4 (32+64 B-25 B^2)\right)\sech^2(y)\nonumber\\
&&+96 \alpha B^4 \left(16 - 248 B - 508 B^2 + 75 B^3\right)~\sech^4(y)\nonumber\\
&&-96 \alpha B^4 \left(88 - 436 B - 824 B^2 + 75 B^3\right)~\sech^6(y)\nonumber\\ 
&&+480 \alpha B^4 \left(16 - 44 B - 76 B^2 + 5 B^3\right)~\sech^8(y)\,,
\eea
Now, we choose
\begin{equation}
\label{alpha}
\alpha=-\frac{1}{64B^4 (32+64 B-25 B^2)}\,,
\end{equation}
which allows to write Eq. (\ref{eqant}) as
\bea\label{dphi2}
\phi^{\prime2}&=&a_1 ~\sech^4(y)-a_2~\sech^6(y)+a_3~\sech^8(y)\,,
\eea
where
\bea
a_1&=&-\frac{3(16 - 248 B - 508 B^2 + 75 B^3)}{2 (32+64 B-25 B^2)}\,,\\
a_2&=&-\frac{3(88 - 436 B - 824 B^2 + 75 B^3)}{2 (32+64 B-25 B^2)}\,,\\
a_3&=&-\frac{15(16 - 44 B - 76 B^2 + 5 B^3)}{2 (32+64 B-25 B^2)}\,.
\eea
Let us rewrite Eq. (\ref{dphi2}) as
\bea
\label{eq26}
\phi^{\prime2}&=&a_3~\sech^4(y)\Big[\frac{a_1}{a_3}-\frac{a_2}{a_3}~\sech^2(y)+\sech^4(y)\Big]\,.
\eea
Now, we define the constants 
\begin{equation}
\frac{a_1}{a_3}=r^2\,;\;\;\;\;
\frac{a_2}{a_3}=2r\,,
\end{equation}
with the restriction
\begin{equation}
\pm\sqrt{\frac{a_1}{a_3}}=\frac{a_2}{2a_3}\,,
\end{equation}
 that is,
\begin{equation}
\label{eq30}
\pm \sqrt{\frac{16 - 248 B - 508 B^2 + 75 B^3}{5(16 - 44 B - 76 B^2 + 5 B^3)}}=\frac{88 - 436 B - 824 B^2 + 75 B^3}{10(16 - 44 B - 76 B^2 + 5 B^3)}\,.
\end{equation}

\begin{figure}[t]
\begin{center}
\includegraphics[scale=0.5]{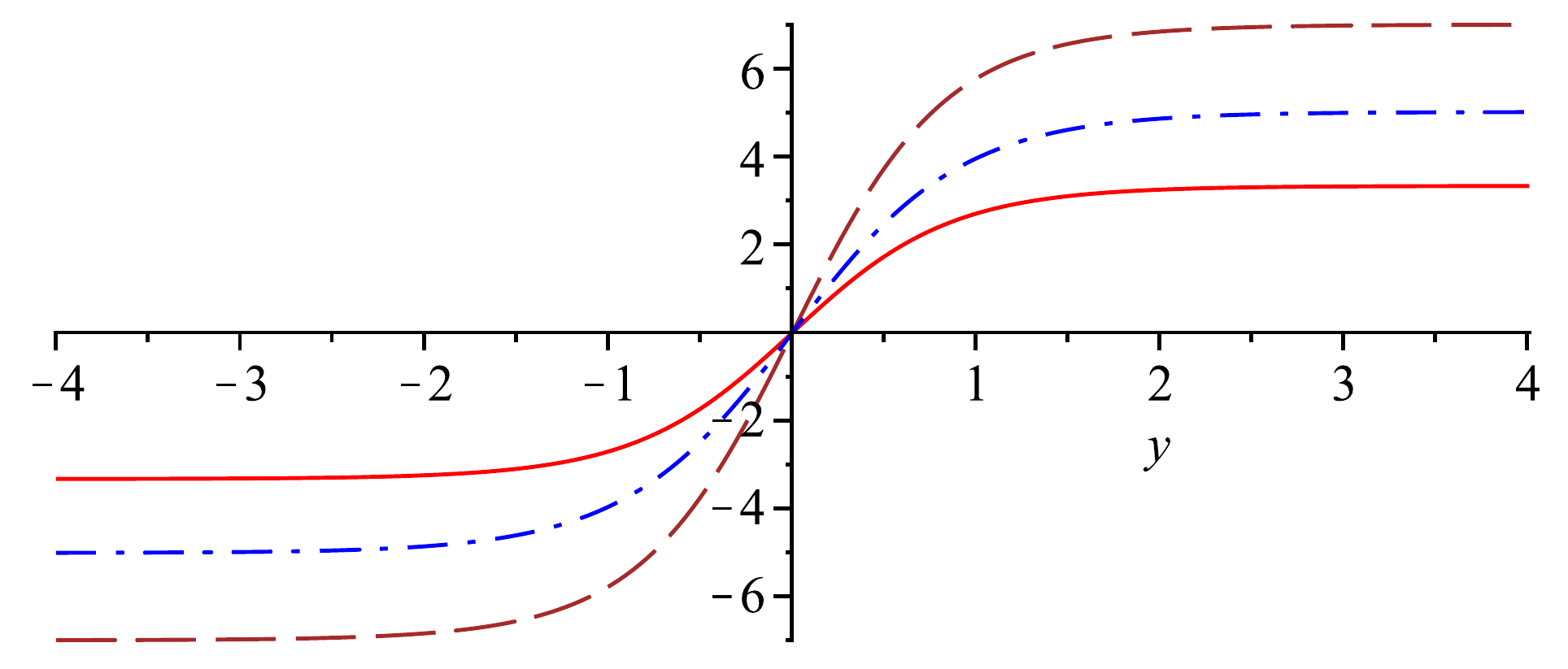}
\end{center}
\vspace{-0.5cm}
\caption{\small{The solutions (\ref{n2sol1}) (red, solid line),  $1/10\times$(\ref{n2sol2}) (brown, dashed line), and $1/2\times$(\ref{n2sol3}) (blue, dotted-dashed line).\label{fig2}}}
\end{figure}

One can show that the possible solutions correspond only to the negative sign in this equation which possesses six solutions, with two of them being essentially complex. From the remaining four solutions, only three satisfy $\phi^{\prime\,2}>0$, they are: $B_1=0.4524$, $B_2=2.9561$ and $B_3=20.0107$. Therefore, if Eq. (\ref{eq30}) is satisfied, from Eq. (\ref{eq26}) we have
\bea
\label{eq32}
\phi^{\prime}=\pm\sqrt{a_3}~\sech^2(y)\Big[r-\sech^2(y)\Big]\,,
\eea
which has the general solution 
\begin{equation}
\phi(y)=\pm\frac13 \sqrt{a_3}~\Big[3r-2-\sech^2(y)\Big]\tanh(y)\,.
\end{equation}
From that, for permitted values of $B$, we obtain
\bea
\phi_1(y)&=&\pm \left(0.5325 \tanh^3(y)- 3.8757 \tanh(y)\right) \mbox{\,\,\,\,\,\,\,\,\,\,\,     for  \,\,\,\,\,\,\,} B=B_1\,,\label{n2sol1}\\
\phi_2(y)&=&\pm \left(14.0826 \tanh^3(y)- 84.3242 \tanh(y)\right) \mbox{\,\,\,\,\,     for  \,\,\,\,\,\,\,} B=B_2\,,\label{n2sol2}\\
\phi_3(y)&=&\pm \left(0.9165 \tanh^3(y)- 10.9778 \tanh(y)\right) \mbox{\,\,\,\,\,\,\,\,     for  \,\,\,\,\,\,\,} B=B_3\,,\label{n2sol3}\
\eea
which are depicted in Fig.~(\ref{fig2}). We note that although the solutions vary similarly, the amplitude of $\phi_2(y)$ is much greater then it is in the other two cases.

\begin{figure}[t]
\begin{center}
\includegraphics[scale=0.27]{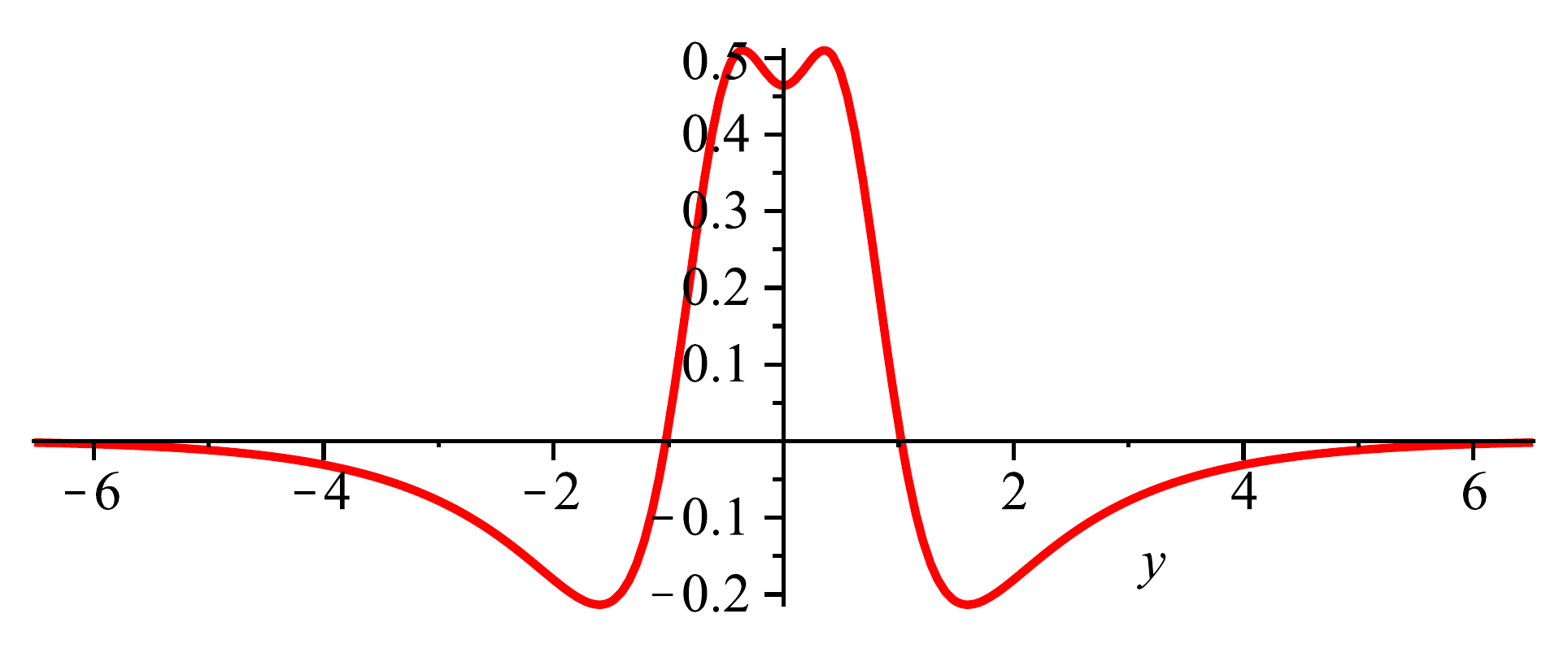}
\includegraphics[scale=0.27]{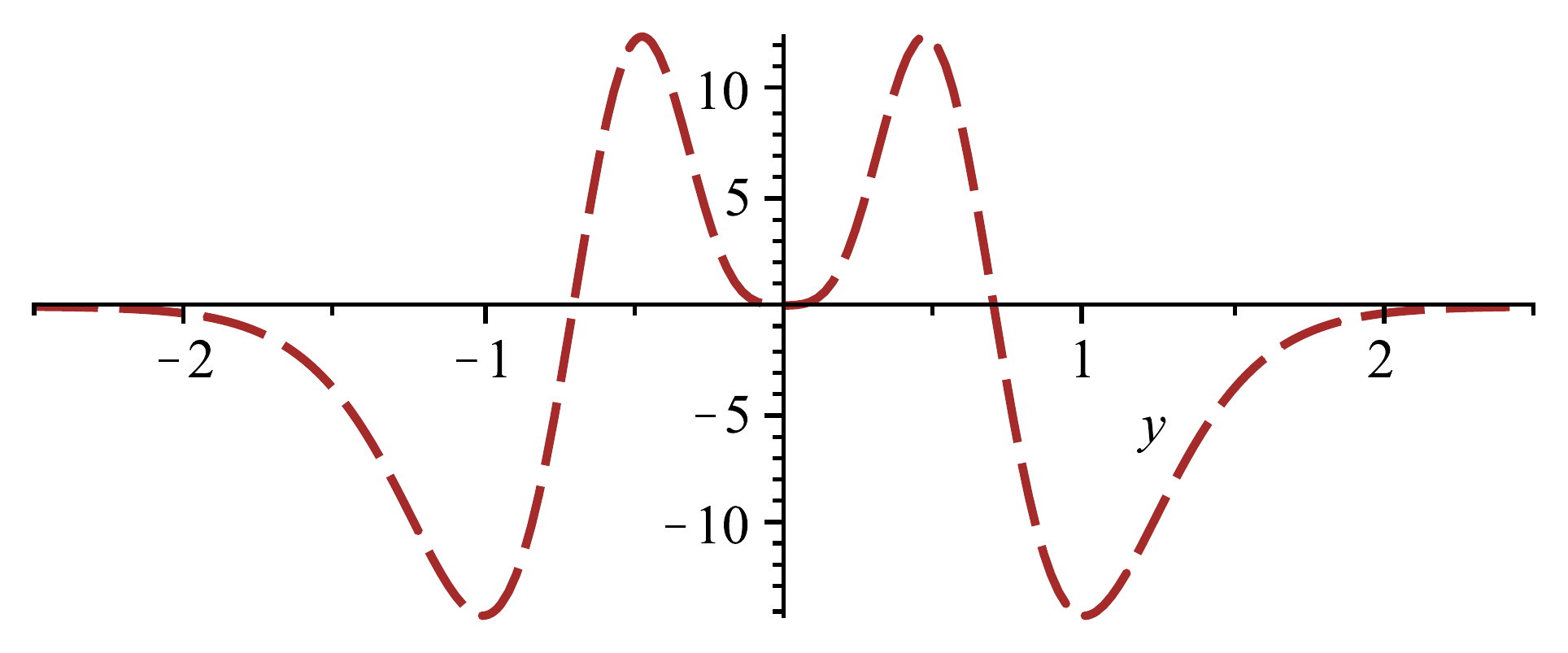}
\includegraphics[scale=0.27]{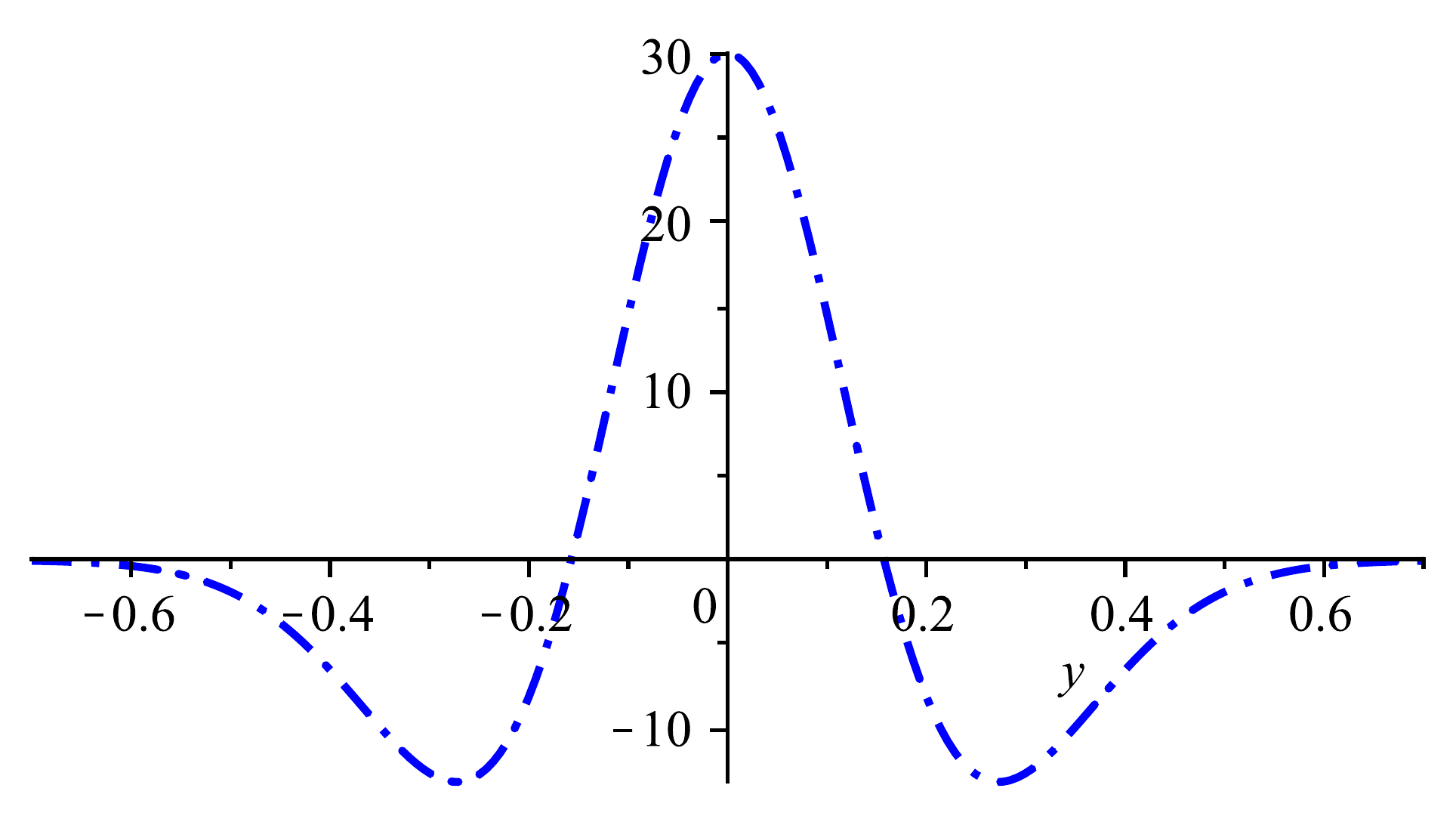}
\end{center}
\vspace{-0.5cm}
	\caption{\small{The energy densities for the solutions (\ref{n2sol1}) (red solid line),  (\ref{n2sol2})(brown dashed line), and (\ref{n2sol3}) (blue dotted-dashed line). \label{fig3}}}
\end{figure}

In Fig.~\ref{fig3} we depict the energy density obtained from (\ref{eneden}), corresponding to each one of the above solutions. We see that the brane has the standard behavior for $B=B_3$, but it splits significantly for $B=B_2$. The splitting  behavior of the brane was studied before in \cite{a,b}, as an effect driven by the source Lagrange density. It was also identified in \cite{blmps}, through modification of the geometry, shown to appear from the inclusion of the $R^2$ term, in the case of non constant curvature. Here it also appear through modification of the geometry, with the inclusion of the $G^2$ term in the Einstein-Hilbert action, for non constant Gauss-Bonnet term. Although the splitting is similar to the case found in Ref.~\cite{blmps}, in the present case it is much more evident, making the core of the brane to behave as it is asymptotically. 

In these cases, the potential can be found explicitly. For the case $n=1$ it is given by
\bea
V(\phi(y))=-\frac32+\frac34\,\sech^2(y)+\frac{9}{4}\sech^4(y)\,.
\eea
The solution for the field is
\begin{equation}
\phi=\sqrt{\frac32}\tanh(y)\Rightarrow y=\mbox{arctanh}\Big[\sqrt{\frac23}\phi\Big] \,.
\end{equation}
Therefore, we can explicitly express the potential $V(\phi)$ as a function of the field:
\be 
V(\phi)=\frac32-\frac72\, \phi^2+\phi^4\,.
\ee
For the case $n=2$ the potential looks like
\bea
V(\phi(y))\!\!\!&=&\!\!\!-3+2160\alpha + \Big(\frac{9}{2}+11664\alpha \Big) \sech^2(y) - 62736\alpha\,\sech^4(y) +\nonumber \\
\!\!\!&&\!\!\!+ 77808 \alpha \,\sech^6(y) - 28512 \alpha \,\sech^8(y)\,.
\eea
where $\alpha$ is given by Eq. (\ref{alpha}).
In this case we have
\begin{equation}
y(\phi)=\mbox{arctanh}(s(\phi))\,,
\end{equation}
where $s(\phi)$ satisfies the algebraic equation
\begin{equation}
a s^3(\phi)-bs(\phi)-\phi=0\,,
\end{equation}
with $a$ and $b$ are the numerical parameters which can be read off from Eqs. (\ref{n2sol1},\ref{n2sol2},\ref{n2sol3}).
Therefore, one has the potential
\be
V(\phi)=\frac32+384 \alpha-\Big(5568 \alpha+\frac92\Big)s^2(\phi)-384 \alpha \,s^4(\phi)+36240 \alpha\, s^6(\phi)-28512\alpha \,s^8(\phi)\,,
\ee
so we conclude that we succeeded to obtain the potential in the form $V=V(\phi)$.

We can go on and use in (\ref{eqphi}) the ansatz $A^{\prime}=Ce^{m\phi}$, $\phi^{\prime}=De^{m\phi}$, as in \cite{frbrane}. We take into account that the equation of motion of the scalar field is the same as in the usual $f(R)$ brane case \cite{frbrane}
\bea
\phi^{\prime\prime}+4A^{\prime}\phi^{\prime}-\frac{dV}{d\phi}=0,
\eea
whose solution is
\bea
V(\phi)=\frac{1}{2m}(mD^2+4CD)e^{2m\phi},
\eea
just as in \cite{frbrane}; so, we find the exponential potential. In this case, the scalar field yields the Liouville-like form:
\be
\phi=-\frac{1}{m}\ln(mD(y-y_0)), \quad\, A=\frac{C}{D}\phi.
\ee
These ansatzes allow to reduce the equation (\ref{eqphi}) to a purely algebraic, although rather complicated, equation on the parameter $m$ which can be solved. Its explicit form is
\bea
(D^2+\frac{3}{2}mCD)e^{2m\phi}&=&-e^{4nm\phi}[24C^3(5C+4mD)]^{n-1}\times\\&\times&
\alpha[6nmDC^3+2n(n-1)[22m^2C^2D^2-6m^3CD^3-2mC^3D]].\nonumber
\eea
We see that here one meets two situations: first, one can have $n=1/2$ (which is rather exotic -- it means that the Gauss-Bonnet term enters the action with a fractional degree), second, the r.h.s. and the l.h.s. of this equation are separately equal to zero. In both cases one rests with some algebraic equations for coefficients $C$ and $D$, arriving at the final expression
$D=-3mC/2$, with $97m^2(n-1)=6$; thus, for the given $n$, we get a fractional $m$.

In summary, in this work we considered the Gauss-Bonnet braneworld model. For this theory, we used the first-order formalism which allowed us to find some exact solutions. First, we succeeded to find the scalar field and its potential in an implicit form, in terms of some auxiliary variable, for the case of the constant Gauss-Bonnet term. Second, we abandoned the condition that the Gauss-Bonnet term is constant, and in this case we found the solutions for the specific forms of the function $f(G)$, that is, $f(G)=aG^n$ with $n=1$ and $n=2$. Also, we found the solution corresponding to the exponential potential. All these solutions are not compatible with the constant scalar curvature, therefore, the anti-de-Sitter space is ruled out, and the supersymmetric extension of the theory seems to be impossible.

As a particularly interesting behavior, we noted the splitting of the brane for non constant Gauss-Bonnet term in the case of $n=2$, that is, for $f(G)=a G^2$; see Fig.~\ref{fig2}. This behavior appears from modification of the geometry, and it is similar to the splitting found before in \cite{blmps}, but here it is much more evident for $B=B_2=2.9561$, with the core of the solution behaving as
it is in the region far away from the brane.

An interesting study concerns stability of the braneworld scenarios that we investigated above. We can follow two distinct approaches, one investigating how the tensorial fluctuations in the metric behave, and the other, adding small parameters to control the behavior of the extra terms
one has introduced in the current study. These and other similar issues are presently under consideration, and we hope to report on them in the near future.

{\bf Acknowledgements.} This work was partially supported by CNPq. The work by A. Yu. P. has been supported by the
CNPq project No. 303438/2012-6.

\end{document}